\documentclass[%
 reprint,
 amsmath,amssymb,
 aps,
]{revtex4-2}

\usepackage{graphicx}% Include figure files
\usepackage{dcolumn}% Align table columns on decimal point
\usepackage{bm}% bold math
\usepackage{tabularx}
\usepackage{subcaption}
\usepackage{xcolor}
\usepackage{svg}

\newcommand{\R}{\mathbb{R}}
\newcommand{\Z}{\mathbb{Z}}

\begin{document}

\allowdisplaybreaks

%\preprint{APS/123-QED}

\title{Delayed Coupling Restores Ising Phase Dynamics in Physical Oscillator Networks}% Force line breaks with \\

\author{Yi Cheng}
  \email{zss7gw@virginia.edu}
 \altaffiliation{Charles L. Brown Department of Electrical and Computer Engineering, University of Virginia, Charlottesville, Virginia 22904, USA.}%Lines break automatically or can be forced with \\
 
\author{Liangtao Dai}
  \email{kvf4sf@virginia.edu}
 \altaffiliation{Charles L. Brown Department of Electrical and Computer Engineering, University of Virginia, Charlottesville, Virginia 22904, USA.}

\author{Mircea R Stan}
  \email{mrs8n@virginia.edu}
 \altaffiliation{Charles L. Brown Department of Electrical and Computer Engineering, University of Virginia, Charlottesville, Virginia 22904, USA.}
 
\author{Zongli Lin}
 \email{zl5y@virginia.edu}
\altaffiliation{Corresponding Author;
Charles L. Brown Department of Electrical and Computer Engineering, University of Virginia, Charlottesville, Virginia 22904, USA.
}

% \collaboration{MUSO Collaboration}%\noaffiliation

%  \homepage{http://www.Second.institution.edu/~Charlie.Author}
% \affiliation{
%  Second institution and/or address\\
%  This line break forced% with \\
% }%
% \affiliation{
%  Third institution, the second for Charlie Author
% \affiliation{%
%  Authors' institution and/or address\\
%  This line break forced with \textbackslash\textbackslash
% }%

% \collaboration{CLEO Collaboration}%\noaffiliation


\begin{abstract}
Oscillator-based Ising machines, in which the phases of coupled self-sustaining oscillators evolve toward decreasing an Ising Hamiltonian, are commonly interpreted as physical realizations of the Ising model. This interpretation, however, requires the phase dynamics generated by the physical oscillator network to match a prescribed Ising dynamics. Here we show that this correspondence is generally not guaranteed. For arbitrary self-sustaining oscillators under weak coupling, we derive the physical phase interaction from the harmonic overlap between the injected waveform and the perturbation projection vector (also referred to as impulse sensitivity function). We find that uncompensated harmonic phase mismatches between these two quantities generate even components in the physical coupling function, causing a network with the correct coupling topology to implement a non-Ising dynamics. We further show that delayed coupling provides a universal phase-compensation mechanism. For a fixed delay, we derive a condition on the delay under which the even components is minimized in the sense of $L^2$-norm, and oscillator examples confirm that the predicted delay substantially suppresses the even components and brings the realized coupling function closer to the prescribed odd interaction. We then show that a periodically modulated delay can, under suitable moment conditions, eliminate the even components in the phase dynamics. These results establish a general design principle for implementing prescribed energy-based dynamics in physical oscillator networks.

\end{abstract}

\maketitle

%\tableofcontents

% \section{\label{sec:level1}First-level heading:\protect\\ The line
% break was forced \lowercase{via} \textbackslash\textbackslash}

\emph{Introduction}---
The phase description of a physical oscillator under injection dates back to Adler’s study of locking phenomena \cite{adler2006study}, which reduced the dynamics of an oscillator driven by a near-resonant periodic signal to a scalar differential equation for its phase difference from the injected signal. This equation provided a compact description of injection locking and pulling, and established the idea that the essential effect of weak periodic perturbations on a self-sustaining oscillator can be captured at the phase level. Decades later, Razavi \cite{razavi2004study} revisited injection locking and pulling from the perspective of electronic oscillator circuits, clarifying how Adler-type phase dynamics arise in practical circuit implementations. More recently, the generalized Adler equation \cite{bhansali2009gen} extended this phase-domain viewpoint beyond ideal sinusoidal injection and specific oscillator topologies by using the perturbation projection vector (PPV), or equivalently the impulse sensitivity function, to describe the phase response of arbitrary oscillators to arbitrary periodic inputs. These developments form a circuit-level foundation for deriving, rather than postulating, the phase interactions generated by physical oscillator circuits.

While Adler-type theories provide a circuit-level phase description for a single oscillator under injection, the Kuramoto model \cite{kuramoto1975international} provides an abstract network-level description of mutually coupled oscillators. Its classical all-to-all sinusoidal form captures collective synchronization, and its later generalizations include weighted interactions and more general coupling functions \cite{acebron2005kuramoto}. In such formulations, however, the pairwise phase interaction is prescribed at the model level rather than derived from the waveform and response properties of a particular physical oscillator.
Oscillator-based Ising machines (OIMs) connect these two viewpoints by seeking a physical realization of prescribed phase-network dynamics using coupled self-sustaining oscillators. In the works  \cite{wang2017oscillator,wang2019oim,wang2021solving}, the phase response of each oscillator to periodic inputs was modeled using a generalized Adler-type equation, while the interactions among many oscillators were cast into a Kuramoto-type network form. The computational interpretation arises by combining these pairwise phase interactions with subharmonic injection locking, which restricts each oscillator to two stable phase states. These two phases encode Ising spins, and the resulting phase dynamics are governed by a Lyapunov function that reduces to the Ising Hamiltonian on the binary phase states. In this sense, OIMs are regarded as physical implementations of Ising-model dynamics rather than merely synchronization systems. This framework has motivated extensive subsequent theoretical and experimental studies of oscillator-based computation \cite{chou2019analog,ahmed2021probabilistic,moy20221,lo2023ising,cheng2024control,wu2025fully,cilasun2025coupled,cheng2025impacts,cheng2025all}.

In the ideal phase description of a coupled-oscillator Ising machine, the pairwise interaction can be written more generally as 
\[
\dot{\theta}_i=-\sum_j J_{ij}\Gamma_{\rm ideal}(\theta_i-\theta_j),
\]
where $\theta_i$ is the phase of the oscillator $i$, $J_{ij}=J_{ji}$ is the coupling weight between oscillators $i$ and $j$, and $\Gamma_{\rm ideal}(\cdot)$ denotes a prescribed coupling function in the ideal class of $2\pi$-periodic odd functions, satisfying $\Gamma_{\rm ideal}(x)=-\Gamma_{\rm ideal}(-x)$ and $\Gamma_{\rm ideal}(x+2\pi)=\Gamma_{\rm ideal}(x)$ \cite{wang2021solving,wang2019oim}. Such a coupling function admits an energy (Lyapunov) function \[E_{\rm ideal}(\theta)=\sum_{i<j}J_{ij}F(\theta_i-\theta_j),\ F(x)=\int_0^x\Gamma_{\rm ideal}(u){\rm d}u,\]  so that the network evolves as a gradient flow of a continuous pairwise energy. This continuous energy function can represent an Ising Hamiltonian when the phase-locking mechanism selects two admissible zeros of $\Gamma_{\rm ideal}(x)$, for example $x=0$ and a nonzero root $x=x_1\in(0,2\pi)$, whose energy values define the two Ising interaction levels. The sinusoidal interaction used in the Kuramoto model, $\Gamma_{\rm ideal}(x)=\sin x$, is the special case in which the two relative phases are $0$ and $\pi$. A physical oscillator network, however, does not directly implement an arbitrarily prescribed $\Gamma_{\rm ideal}(\cdot)$. Instead, its phase dynamics take the form
\[
\dot{\theta}_i=-\sum_j J_{ij}{\Gamma}_{\rm phys}(\theta_i-\theta_j),
\]
where ${\Gamma}_{\rm phys}(\cdot)$ is the coupling function  generated by the waveform, injection port, and phase response of the oscillator. In general, the physically realized coupling function need not belong to the ideal class. In particular, ${\Gamma}_{\rm phys}(\cdot)$ may fail to satisfy the required odd symmetry, so that ${\Gamma}_{\rm phys}(\cdot)\neq \Gamma_{\rm ideal}(\cdot)$ for any admissible ideal coupling function $\Gamma_{\rm ideal}(\cdot)$. A physical oscillator network with the correct coupling matrix $J_{ij}$ can therefore realize phase dynamics that are not equivalent to the intended Ising dynamics.

In this work, we address this physical-realizability problem for general self-sustaining oscillator networks under weak coupling. We derive the physically realized coupling function \({\Gamma}_{\rm phys}(\cdot)\) from the oscillator waveform and the PPV at the injection port, and show that the deviation from the prescribed Ising dynamics is governed by harmonic phase mismatches between these two quantities. This analysis reveals that delayed coupling is a phase-compensation mechanism that can systematically reduce the mismatch between the physical interaction and the prescribed odd Ising interaction. For fixed coupling delay, we derive an optimal-delay condition that minimizes the non-odd component of the physical coupling function in the sense of \(L^2\)-norm, and demonstrate its effectiveness using several physical oscillator examples. We further show that periodically modulated coupling delay can, under suitable moment conditions, eliminate the even component of the averaged physical interaction. These results establish a design principle for restoring Ising-like phase dynamics in physical oscillator machines and, more generally, for implementing prescribed energy-based dynamics in physical oscillator networks.

% In this work, we address this physical-realizability problem for general self-sustaining oscillator networks under weak coupling. We derive the physically realized coupling function $\widetilde{\Gamma}_{\rm phys}(\cdot)$ from the oscillator waveform and the PPV at the injection port, and show that the deviation from the prescribed Ising dynamics is governed by harmonic phase mismatches between these two quantities. This analysis reveals that delayed coupling is a phase-compensation mechanism that can systematically reduce the mismatch between $\widetilde{\Gamma}_{\rm phys}(\cdot)$ and $\Gamma_{\rm ideal}(\cdot)$. We derive an expression for the optimal delay and demonstrate its effectiveness using several physical oscillator examples. These results establish a design principle for restoring Ising dynamics in physical oscillator machines and, more generally, for implementing prescribed energy-based dynamics in physical oscillator networks.

\emph{Physical phase interactions}$-$ 
We first derive the phase interaction generated by a weakly coupled network of self-sustaining oscillators. For a single oscillator under a weak perturbation $b(t)$ applied at a given port, the phase-reduced dynamics can be written as \cite{levantino2012computing}
\[
\dot{\alpha}=p(ft+f\alpha)b(t),
\]
where $p(\cdot)$ is the PPV associated with that port, $f$ is the fundamental frequency of the oscillator, and $\alpha$ denotes the slow phase shift. We next extend this description to a network of weakly coupled oscillators. Suppose that the waveform at the selected port of oscillator $i$ is a periodic signal $x_i(f t) = \sum_{l=-\infty}^{+\infty} a_l{\rm e}^{\imath 2\pi f l t}$, with the Fourier coefficient $a_l = |a_l|{\rm e}^{\imath \varphi_l}$, and that the corresponding PPV is expanded as $p(ft) = \sum_{k=-\infty}^{+\infty} p_k{\rm e}^{\imath 2\pi f k t}$, with the Fourier coefficient $p_k = |p_k|{\rm e}^{\imath \phi_k}$. For oscillator $i$, we model the weak perturbation $b_i(t)$ generated by the rest of the network as
$b_i(t) = \sum_{j=1}^N J_{ij} x_j(ft+f\alpha_j).$
Substituting this perturbation into the single-oscillator phase-reduction equation and applying fast-period averaging over one oscillation cycle, we obtain
\begin{eqnarray*}
    \dot \alpha_i(t) \!\!&\simeq&\!\! \sum_{j=1}^N J_{ij}p_0a_0 + 2 \sum_{j=1}^NJ_{ij}\sum_{k=1}^\infty |p_k||a_k| \\
    &&\!\!\times \cos\left(2\pi f k(\alpha_i-\alpha_j)+\phi_k-\varphi_k\right),
\end{eqnarray*}
where detailed derivation can be found in Section 1 of the Supplementary Material. The first term of the right hand side is independent of the phase differences and represents a DC-induced frequency-bias contribution rather than a pairwise phase interaction. In what follows, we assume a DC-blocked coupling path, equivalently $a_0=0$, so that this term is removed and only the oscillatory harmonic interactions remain. Under this assumption, the averaged equation can be written as, \[\dot{\alpha}_i=-\sum_j J_{ij}{\Gamma}_{\rm phys}(\alpha_i-\alpha_j),\] with 
\[{\Gamma}_{\rm phys}(x) = -2 \sum_{k=1}^\infty |p_k||a_k|\cos\left(2\pi f k x+\phi_k-\varphi_k\right).\]
Here $|a_k|$ and $\varphi_k$ are respectively the magnitude and phase of the $k$-th Fourier coefficient $a_k$ of the oscillator waveform at the selected port, while $|p_k|$ and $\phi_k$ are the corresponding magnitude and phase of the $k$-th Fourier coefficient $p_k$ of the PPV at the same port. Thus, the physically realized coupling function is fixed by the harmonic overlap between the oscillator waveform and the PPV, rather than prescribed independently at the phase-model level. 

\emph{Dynamical mismatch}---The ideal Ising dynamics require the pairwise coupling function to belong to the class of $2\pi$-periodic odd functions. The physically realized coupling function derived above, however, does not generally satisfy this condition. Let \[\delta_k = \phi_k-\varphi_k\] denote the phase difference between the $k$-th harmonic of the PPV and the $k$-th harmonic of the oscillator waveform. Then,
\[{\Gamma}_{\rm phys}(x) = -2 \sum_{k=1}^\infty |p_k||a_k|\cos\left(2\pi f k x+\delta_k\right).\]
Using \[\cos(2\pi f k x+ \delta_k)\!\! = \!\cos(\delta_k)\cos(2\pi f k x)\!-\sin(\delta_k)\sin(2\pi f k x),\]
we obtain the decomposition
\begin{eqnarray*}
    {\Gamma}_{\rm phys}(x) \!\!&=&\!\! 2\sum_{k=1}^\infty |p_k||a_k|\sin(\delta_k)\sin(2\pi f k x) \\
    &&\!\!-2\sum_{k=1}^\infty |p_k||a_k|\cos(\delta_k)\cos(2\pi f k x).
\end{eqnarray*}
The first term is odd in the phase difference $x$, whereas the second term is even. Therefore, ${\Gamma}_{\rm phys}(\cdot)$ belongs to the ideal class only if the even component vanishes, which requires \[\cos(\delta_k) = 0\] 
for every harmonic with nonzero $|p_k||a_k|$. In a physical oscillator, this condition is not generally guaranteed because $\delta_k$ is fixed by the harmonic phase relation between the PPV and the oscillator waveform at the selected port. Thus, the mismatch between ${\Gamma}_{\rm phys}(\cdot)$ and $\Gamma_{\rm ideal}(\cdot)$ originates from uncompensated waveform-PPV phase offsets, rather than from the coupling matrix itself.

This observation suggests a direct compensation strategy. A delay in the coupling path shifts the phases of the waveform harmonics and therefore changes the effective values of $\delta_k$. By choosing this delay appropriately, one can suppress the even component of  ${\Gamma}_{\rm phys}(\cdot)$ and make the realized physical interaction closer to the prescribed odd Ising interaction.

\emph{Compensation by fixed coupling delay}--- With a coupling delay $\tau$, the waveform injected from oscillator $j$ becomes 
\[x_j(f(t+\alpha_j-\tau)) = \sum_{l=-\infty}^\infty a_l {\rm e}^{\imath 2\pi f l (t+\alpha_j-\tau)}.\]
Repeating the averaging calculation, as described in detail in Section 2 of the Supplementary Material, gives
\[\dot \alpha_i(t) \simeq -\sum_{j=1}^N J_{ij} {\Gamma}_{\rm phys, \tau}(\alpha_i-\alpha_j),\]
where
\[{\Gamma}_{\rm phys, \tau}(x) \\= -2\sum_{k=1}^\infty |p_k||a_k|\cos(2\pi f kx+\delta_k +2\pi f k\tau).\]
Thus, the delay shifts the phase of the $k$-th harmonic by $2\pi f k \tau$. This provides a direct way to compensate the harmonic phase mismatch between the waveform and the PPV.

To choose the delay, we seek to suppress the even component of the delayed coupling function while enhancing its odd component. Introducing the phase difference $\theta = 2\pi f x$, the even part of ${\Gamma}_{\rm phys, \tau}(x)$ is 
\[g_{\tau}(\theta) = -2\sum_{k=1}^\infty |p_k||a_k| \cos(\delta_k+2\pi f k\tau)\cos(k\theta),\]
and the odd part is 
\[h_{\tau}(\theta) = 2\sum_{k=1}^\infty |p_k||a_k| \sin(\delta_k+2\pi f k\tau)\sin(k\theta),\]
We define the following two objectives in the form of $L^2$-norm over $[-\pi,\pi)$,
\begin{eqnarray}
    G(\tau) &=& \int_{-\pi}^{\pi} g_{\tau}^2(\theta) {\rm d}\theta, \label{obj1} \\ 
     H(\tau) &=& \int_{-\pi}^{\pi} h_{\tau}^2(\theta) {\rm d}\theta. \label{obj2}
\end{eqnarray}
By orthogonality of the Fourier basis functions,
\[G(\tau) = 4\pi \sum_{k=1}^\infty |p_k|^2|a_k|^2\cos^2(\delta_k+2\pi f k \tau), \]
and 
\[H(\tau) = 4\pi \sum_{k=1}^\infty |p_k|^2|a_k|^2\sin^2(\delta_k+2\pi f k \tau).\]
Since $G(\tau)+H(\tau) = 4\pi \sum_{k=1}^\infty |p_k|^2|a_k|^2$ is independent of $\tau$, the two design objectives are not competing within this constant-delay family. Minimizing the even component $G(\tau)$  therefore also maximizes the odd component $H(\tau)$.
The optimal compensation delay is therefore
\[\tau^\star = \arg \min_{\tau \in [0,1/f)}G(\tau) = \arg \max_{\tau \in [0,1/f)}H(\tau) .\]
If $\tau^\star$ is an interior local minimizer, it satisfies
\[\sum_{k=1}^\infty k|p_k|^2|a_k|^2\sin(2\delta_k+4\pi f k \tau^\star) = 0.\]
Equivalently, writing $\theta^\star = 4\pi f \tau^\star$, the candidate delays are obtained from the roots of
\[\sum_{k=1}^\infty k|p_k|^2|a_k|^2\sin(2\delta_k+k\theta) = 0.\]
It is then shown in Section 2 of the Supplementary Material that
\[\tau^\star = \frac{\theta^\star}{4\pi f} + \frac{n}{2f},\]
with $n\in \Z$ chosen so that $\tau^\star \in [0,1/f)$. The final delay is selected by evaluating $G(\tau)$ over all candidate solutions. At $\tau = \tau^\star$,  the even component of ${\Gamma}_{\rm phys, \tau}(\cdot)$ is minimized while its odd component is maximized in the $L^2$ sense. Within the one-parameter family generated by a constant coupling delay, this choice gives the physical interaction with the smallest non-odd contribution, thereby minimizing the dynamical mismatch associated with the violation of the prescribed odd Ising interaction. 

\begin{figure}[htbp!]
    \centering
    \includegraphics[width=1.0\linewidth]{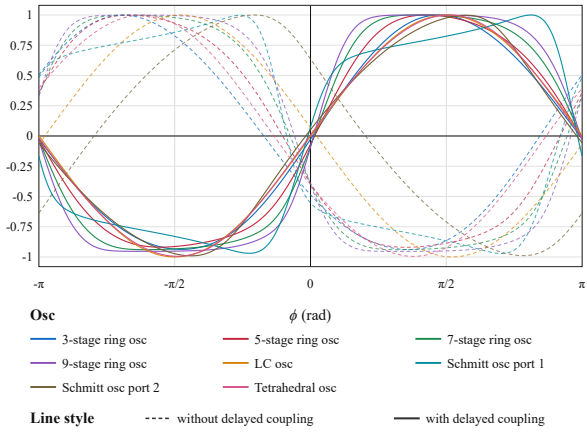}
    \caption{Coupling functions of oscillators with and without the optimal delayed coupling on $3$-stage ring oscillator, $5$-stage ring oscillator, $7$-stage ring oscillator, $9$-stage ring oscillator \cite{hajimiri1999jitter}, LC oscillator \cite{hegazi2001filtering},  Schmitt oscillator port 1,  Schmitt oscillator port 2 \cite{schmitt1938thermionic}, and tetrahedral oscillator \cite{smith2021analysis}.}
    \label{coupling_functions_normalized}
\end{figure}

\begin{figure}[htbp!]
    \centering
    \includegraphics[width=1\linewidth]{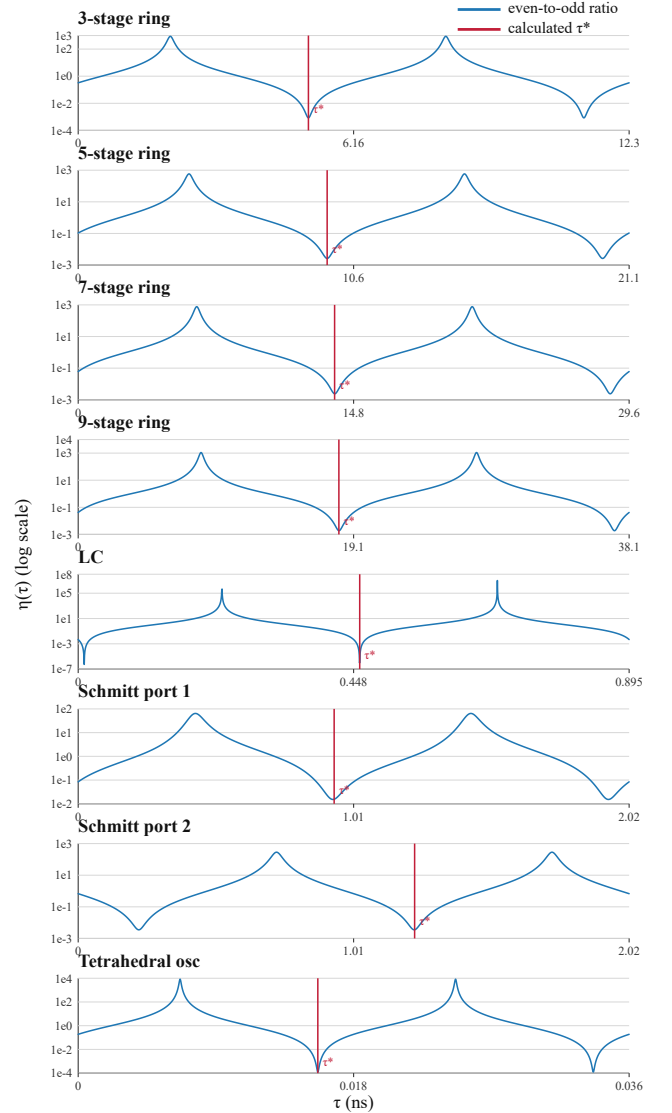}
    \caption{Even-to-odd ratio over one oscillation period $1/f$ for different oscillators with the red point as predicted optimal delay $\tau^\star$. }
    \label{GH_tau_sweep}
\end{figure}

The delayed physical coupling function can be measured directly from the oscillator waveform and PPV \cite{bhansali2009gen,wu2025fully}, which is computed as
\[{\Gamma}_{\rm phys, \tau}(\phi) = -\frac{1}{2\pi}\int_{0}^{2\pi} p(\phi+\varphi)x_j(\varphi-2\pi f \tau){\rm d}\varphi,\]
where $x_j(\cdot)$ is the waveform of the injecting oscillator at the selected port. The case $\tau=0$ corresponds to the physical coupling function without delayed coupling, while $\tau=\tau^\star$ gives the optimized delayed interaction. We evaluated this coupling function for seven oscillator implementations. 
Fig.~\ref{coupling_functions_normalized} shows the coupling functions with and without optimal delayed coupling.
To quantify the effect of the coupling delay on the realized coupling function, we use the even-to-odd energy ratio
\[\eta(\tau) = \frac{G(\tau)}{H(\tau)}\]
as a measure of the non-odd contribution. For coupling function ${\Gamma}_{\rm phys,\tau}(\cdot)$ computed from the measured oscillator waveform and PPV, we decompose it into its even and odd components,
\[g_{\tau}(\phi) = \frac{{\Gamma}_{\rm phys,\tau}(\phi)+{\Gamma}_{\rm phys,\tau}(-\phi)}{2},\]
and 
\[h_{\tau}(\phi) = \frac{{\Gamma}_{\rm phys,\tau}(\phi)-{\Gamma}_{\rm phys,\tau}(-\phi)}{2}.\]
The corresponding energies $G(\tau)$ and $H(\tau)$ are then computed from Eqs. \eqref{obj1} and \eqref{obj2}. Fig.~\ref{GH_tau_sweep} shows the variation of $\eta(\tau)$ over one oscillation period $1/f$ for the seven oscillators, together with the theoretically predicted optimal delay $\tau^\star$. The minima of the experimentally evaluated $\eta(\tau)$ agree well with the predicted $\tau^\star$, confirming that the proposed delay compensation effectively suppresses the even component of the physical coupling function.

We next examine how coupling delay affects the Max-Cut performance of oscillator networks. We consider \(3\times 3\) networks composed of either three-stage ring oscillators or LC oscillators, with four representative coupling topologies: the mesh, king, queen, and all-to-all graphs. For each topology and each value of the coupling delay \(\tau\), we perform repeated optimization runs and record the maximum, minimum, and mean cut values obtained by the network.

\begin{figure*}[t]%[htbp!]
    \centering
    \includegraphics[width=0.75\linewidth]{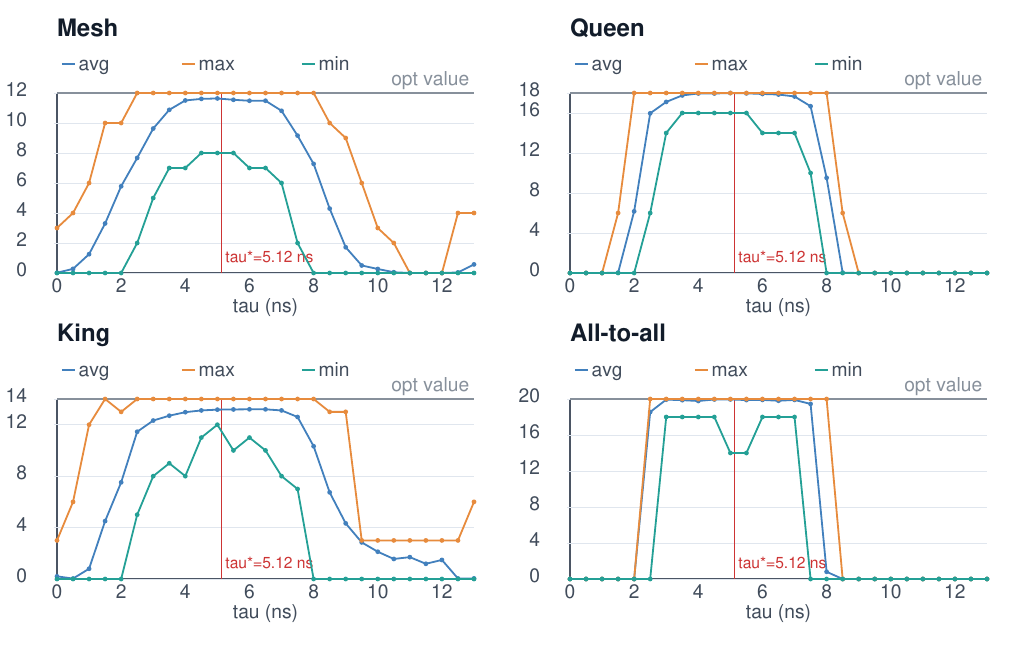}\vspace{-20pt}
    \caption{Max-Cut tests as function of $\tau$ for a three-stage ring oscillator network}
    \label{Ring_tau}
\end{figure*}

\begin{figure*}%[h!]%[htbp!]
    \centering
    \includegraphics[width=0.75\linewidth]{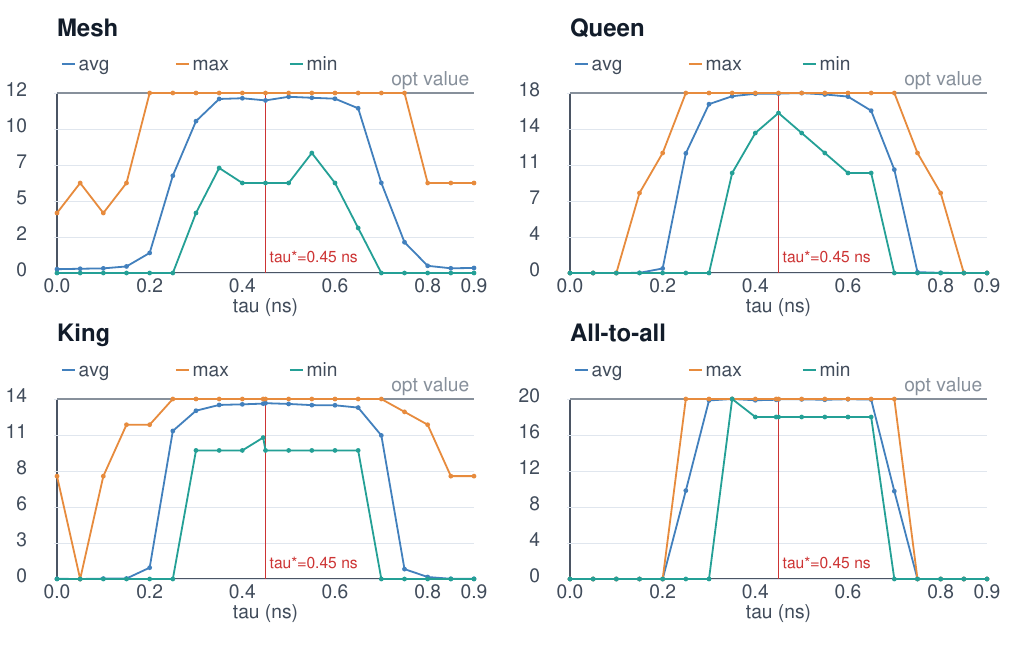}\vspace{-20pt}
    \caption{Max-Cut tests as function of $\tau$ for an LC oscillator network}
    \label{LC_tau}
\end{figure*}

Figs.~\ref{Ring_tau} and \ref{LC_tau} show the Max-Cut results as functions of \(\tau\) for the ring-oscillator and LC-oscillator networks, respectively. The theoretically predicted optimal delay \(\tau^\star\) is marked in each panel. Across all four graph topologies and for both oscillator implementations, the maximum, minimum, and average cut values exhibit a broad high-performance region centered approximately around \(\tau^\star\). Outside this region, all three performance measures decrease substantially. These results show that the delay predicted by the phase-interaction analysis not only suppresses the non-odd component of the physical coupling function, but also improves the reliability and solution quality of oscillator networks solving Max-Cut problems. The detailed experimental setup and procedure are provided in  Section 4 of the the Supplementary Material.

\emph{Compensation by periodically modulated coupling delay}---  An optimized fixed delay can reduce the even component of the physical coupling function, but a single constant delay does not generally eliminate the phase mismatch of all contributing harmonics simultaneously. We therefore consider a periodically modulated coupling delay. Let \(T_0=1/f\) denote the carrier period, and let the delay satisfy \[
\tau(t+T_{\rm delay})=\tau(t),
\]
where \(T_{\rm delay}\) is the delay-modulation period. Define the dimensionless oscillator phase shifts and delay phase by
\[
\vartheta_i(t)=2\pi f\alpha_i(t),
\qquad
\varrho(t)=2\pi f\tau(t),
\]
respectively. We further characterize their variation by
\[
T_\alpha^{-1}=\max_{i,t}|\dot{\vartheta}_i(t)|,
\qquad
T_\varrho^{-1}=\max_t|\dot{\varrho}(t)|.
\]
Under time-scale separation,
\[
T_0\ll T_{\rm delay}\ll T_\alpha,
\qquad
T_0\ll T_\varrho,
\]
the phase dynamics can be averaged successively over the carrier period and over one delay-modulation period. The resulting averaged slow phase dynamics, as derived in Section 3 of the Supplementary Material, are given by
\[\dot \alpha_i(t) \simeq -\sum_{j=1}^N J_{ij} {\Gamma}_{\rm phys, eff}(\alpha_i-\alpha_j),\]
where 
\begin{eqnarray*}
    {\Gamma}_{\rm phys, eff}(x) \!\!&=&\!\! -2\sum_{k=1}^\infty |p_k||a_k|\\
&&\!\!\times\left(C_k\cos(2\pi fk x)-S_k\sin(2\pi f k x)\right),
\end{eqnarray*}
with 
\[C_k(\tau) = \frac{1}{T_{\rm delay}}\int_{0}^{T_{\rm delay}}\cos(\delta_k+2\pi f k \tau(t)){\rm d}t,\]
and 
\[S_k(\tau) = \frac{1}{T_{\rm delay}}\int_{0}^{T_{\rm delay}}\sin(\delta_k+2\pi f k \tau(t)){\rm d}t.\]
The first averaging step is the same carrier-scale averaging used for a fixed delay, which removes rapidly oscillating terms over one oscillation cycle and yields a phase interaction that depends parametrically on the slowly varying delay phase $\varrho(t)$. The second averaging step is specific to the periodically modulated delay. Since $T_{\rm delay} \ll T_\alpha$, the phase differences are approximately constant during one delay-modulation cycle. The carrier-averaged interaction can therefore be further averaged over $T_{\rm delay}$, yielding an effective coupling function that describes the evolution of the phase differences averaged over one delay-modulation period. This description captures the net phase evolution from one modulation-cycle boundary to the next, rather than the detailed interaction within each modulation cycle.

The periodic delay waveform $\tau(t)$ determines the effective contribution of each harmonic through the complex modulation moments
\[M_k(\tau) = \frac{1}{T_{\rm delay}}\int_{0}^{T_{\rm delay}} {\rm e}^{\imath k \varrho(t)} {\rm d}t, k=1,2,\ldots,\]
where $\varrho(t) = 2\pi f \tau(t)$.
The coefficients of the effective coupling function satisfy
\[C_k(\tau)+\imath S_k(\tau) = {\rm e}^{\imath \delta_k}M_k(\tau).\]
Accordingly, the even contribution of the $k$-th harmonic vanishes when $C_k = {\rm Re}\{{\rm e}^{^{\imath \delta_k}}M_k(\tau)\} = 0$. A sufficient condition on the periodic delay waveform $\tau(t)$ for eliminating the even contribution of every retained harmonic, as derived in Section 3 of the Supplementary Material, is therefore
\[\frac{1}{T_{\rm delay}} \int_{0}^{T_{\rm delay}} {\rm e}^{\imath 2\pi fk \tau(t)}{\rm d}t = \imath \beta_k {\rm e}^{-\imath \delta_k},\ \beta_k\in \R,\]
and hence,
\[C_k(\tau)=0,\qquad S_k(\tau)=\beta_k,\]
which in turn implies that
\[{\Gamma}_{\rm phys, eff}(x) = 2\sum_{k=1}^\infty |p_k||a_k|\beta_k \sin(2\pi f kx),\]
which is odd within the retained harmonic.

Since a single periodic waveform $\tau(t)$ determines all of the moments $M_k(\tau)$ simultaneously, the coefficients $\beta_k$ cannot in general be specified independently. Whether the condition can be satisfied exactly for multiple harmonics depends on the set of modulation moments realizable by the chosen delay waveform.

\emph{Conclusion and outlook}---
In this work, we studied the dynamical mismatch between physical oscillator networks and prescribed Ising phase dynamics. Starting from phase reduction, we showed that the coupling function realized by a weakly coupled network of self-sustaining oscillators is determined by the harmonic overlap between the oscillator waveform and the PPV at the coupling port. This physical coupling function generally contains an even component, which breaks the odd symmetry required by the ideal Ising interaction and leads to a mismatch between the implemented oscillator dynamics and the intended energy-based model.

We showed that coupling delay provides a systematic way to reduce this mismatch. For a constant coupling delay, we derived an optimal delay condition that minimizes the even component of the physical coupling function in the sense  of \(L^2\)-norm while maximizing its odd component within the constant-delay family. Experiments based on extracted oscillator waveforms and PPVs, together with Max-Cut tests on oscillator networks, confirmed that the predicted delay substantially suppresses the non-odd contribution and improves the solution quality of the resulting oscillator dynamics. However, because a single fixed delay shifts all harmonics in a constrained way, it generally reduces but does not completely eliminate the mismatch. To overcome this limitation, we further introduced periodically modulated coupling delay. By averaging over both the carrier oscillation and the delay-modulation period, we derived an effective coupling function that governs the slow phase evolution. We then identified sufficient conditions on the modulation moments of \(\tau(t)\) under which the even component of every retained harmonic vanishes. Under these conditions, the effective coupling function becomes purely odd, so that the oscillator network can realize the prescribed Ising phase interaction at the averaged-dynamics level. These results suggest that delay engineering can serve as a general mechanism for bridging the gap between abstract Ising phase models and their physical implementation in self-sustaining oscillator networks.

Our findings open several directions, including the circuit-level realization of periodically modulated coupling delays under bandwidth, noise, and delay-range constraints; the joint design of oscillator waveforms, PPVs, coupling ports, and delay modulation to synthesize prescribed phase interactions; and the extension of this framework beyond Ising dynamics to more general energy-based models, such as Potts and clock models. More broadly, our results suggest that realizing prescribed energy-based dynamics in physical oscillator networks should be viewed not only as a graph-programming problem, but also as a waveform--response--delay synthesis problem.

% To express the condition for eliminating the even component, define the complex modulation moments
% \[M_k = \frac{1}{T_\tau}\int_{0}^{T_\tau} {\rm e}^{\imath k \varrho(t)} {\rm d}t, k=1,2,\ldots\]
% Then 
% \[C_k+\imath S_k = {\rm e}^{\imath \delta_k}M_k.\]
% Hence, the even component of the $k$-th harmonic vanishes whenever
% \[C_k = {\rm Re}\{{\rm e}^{^{\imath \delta_k}}M_k\} = 0.\]
% A sufficient harmonic design condition is therefore
% \[M_k = \imath \beta_k {\rm e}^{-\imath (\phi_k-\varphi_k)},\  \beta_k \in \R,\]
% for every retained harmonic $k$. Under this condition, $C_k=0$ and  $S_k =\beta_k$, and therefore
% \[\widetilde{\Gamma}_{\rm phys, eff}(x) = 2\sum_{k=1}^\infty |p_k||a_k|\beta_k \sin(2\pi f kx),\]
% which is purely odd. This condition is a sufficient design target. Whether it can be satisfied simultaneously for multiple harmonics depends on the set of modulation moments realizable by a single periodic delay waveform $\tau(t)$.   

\emph{Acknowledgments}$-$ This material is based upon work supported by the National Science Foundation under grant no. 2328961.

% \nocite{*}

\bibliography{Main_Text_refIsing}% Produces the bibliography via BibTeX.

\clearpage
\onecolumngrid
\noindent\includegraphics[page=1,width=\textwidth,height=\textheight,keepaspectratio]{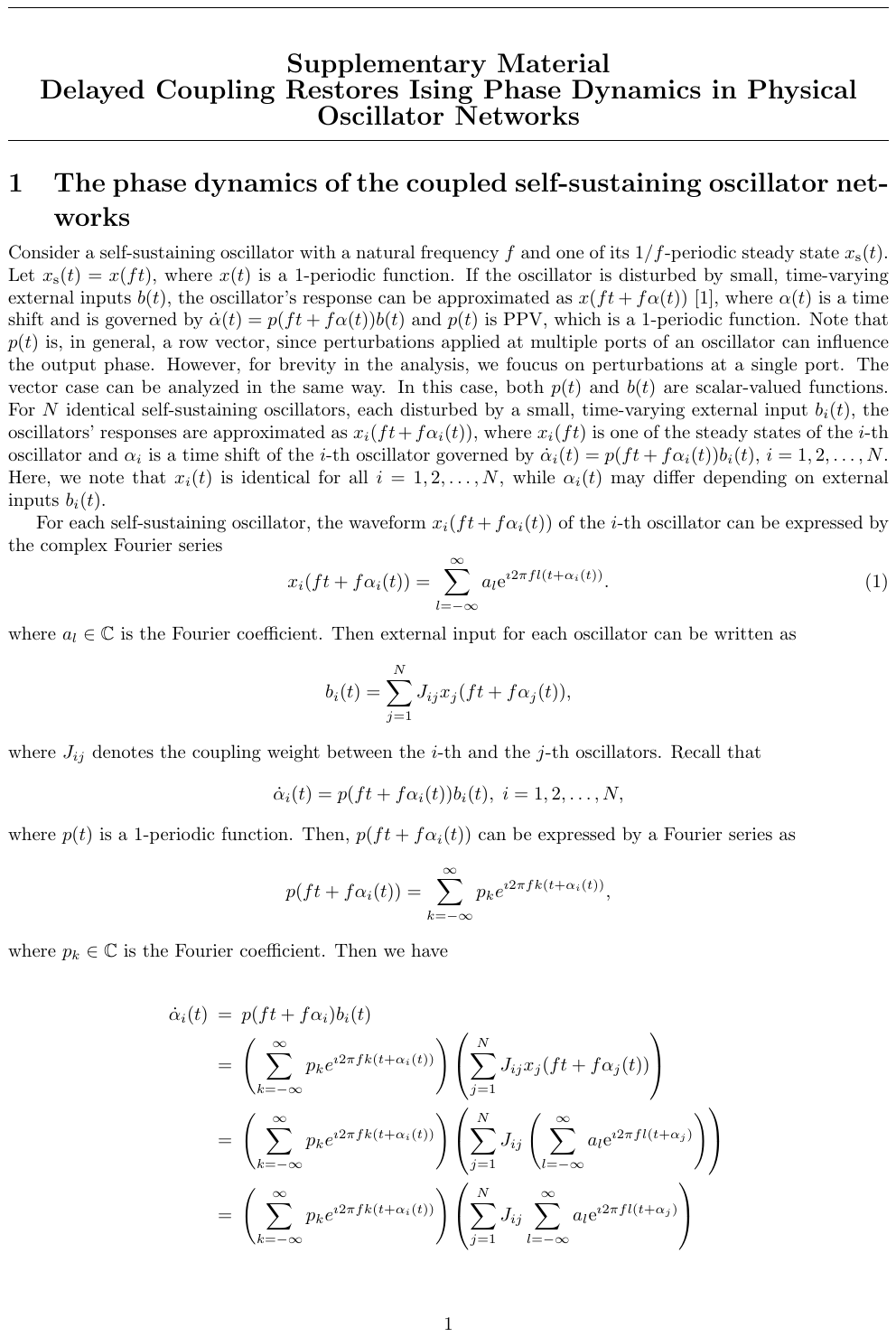}\clearpage
\noindent\includegraphics[page=2,width=\textwidth,height=\textheight,keepaspectratio]{Supplementary.pdf}\clearpage
\noindent\includegraphics[page=3,width=\textwidth,height=\textheight,keepaspectratio]{Supplementary.pdf}\clearpage
\noindent\includegraphics[page=4,width=\textwidth,height=\textheight,keepaspectratio]{Supplementary.pdf}\clearpage
\noindent\includegraphics[page=5,width=\textwidth,height=\textheight,keepaspectratio]{Supplementary.pdf}\clearpage
\noindent\includegraphics[page=6,width=\textwidth,height=\textheight,keepaspectratio]{Supplementary.pdf}\clearpage
\noindent\includegraphics[page=7,width=\textwidth,height=\textheight,keepaspectratio]{Supplementary.pdf}\clearpage
\noindent\includegraphics[page=8,width=\textwidth,height=\textheight,keepaspectratio]{Supplementary.pdf}\clearpage
\noindent\includegraphics[page=9,width=\textwidth,height=\textheight,keepaspectratio]{Supplementary.pdf}\clearpage
\noindent\includegraphics[page=10,width=\textwidth,height=\textheight,keepaspectratio]{Supplementary.pdf}\clearpage
\noindent\includegraphics[page=11,width=\textwidth,height=\textheight,keepaspectratio]{Supplementary.pdf}\clearpage
\noindent\includegraphics[page=12,width=\textwidth,height=\textheight,keepaspectratio]{Supplementary.pdf}\clearpage
\noindent\includegraphics[page=13,width=\textwidth,height=\textheight,keepaspectratio]{Supplementary.pdf}\clearpage
\noindent\includegraphics[page=14,width=\textwidth,height=\textheight,keepaspectratio]{Supplementary.pdf}\clearpage
\noindent\includegraphics[page=15,width=\textwidth,height=\textheight,keepaspectratio]{Supplementary.pdf}\clearpage
\noindent\includegraphics[page=16,width=\textwidth,height=\textheight,keepaspectratio]{Supplementary.pdf}\clearpage
\end{document}